\documentclass[prb,intlimits,twocolumn]{revtex4}
\usepackage{amsmath}
\usepackage{graphicx}

\usepackage{amssymb}
\usepackage{revsymb}
\usepackage{amsfonts}

\begin{document}
\title{Infrared detector based on conduction band intersubband transitions in a heterojunction between two quantum wires.}
\author{V.A. Shuvayev, L.I. Deych, and A.A. Lisyansky}
\affiliation{Department of Physics, Queens College, The City
University of New York, Flushing, NY 11367}
\author{M.J. Potasek}
\affiliation{Departments of Physics and Electrical Engineering,
City College, The City University of New York, New York, NY 10031 and\\
Courant Institute, New York University, New York, N.Y. 10021}

\begin{abstract}
In this paper we study the feasibility of an infrared detector based
on intersubband transitions in the conduction band of the junction
between two semiconductor quantum wires. We show that by varying the
radius of the wires it is possible to engineer a band structure of
the junction that would be favorable for creating and detecting
photocurrent. The suggested concept also allows for broadband
detection based on arrays of wires with different radii.
\end{abstract}
\maketitle
Infrared (IR) photo detectors operating in the medium wavelength IR
between 3-5 $\mu m$, long wavelength IR between 8-12 $\mu m$, and
very long IR, are important for many applications including thermal
imaging and remote sensing. Intersubband transitions are a
typical mechanism for quantum well infrared photodetection%
\cite{QWIP}, and have various attractive features. First, the
position of the subbands depends significantly on the geometric
dimensions of the system, and can be effectively controlled. Second,
phonon scattering is reduced by the quantum confinement, enabling
the detector to operate at higher temperatures. However, quantum
well infrared photodetectors have one significant disadvantage:
intersubband transitions in these systems cannot be excited by light
at normal incidence. Therefore, one needs to use special coupling
mechanisms such as two-dimensional grating couplers. One way to
avoid this problem and simplify the design of the detectors is to
replace quantum wells with quantum wires that do not suffer from
this drawback. Thus, using quantum wires instead of quantum wells,
one can develop more convenient detectors sensitive to normally
incident light. Using the nanopatterned template
technique~\cite{nanopattern}, one can fabricate quantum wires with a
predefined radius and length. Aluminum is anodized in an
electrochemical cell where it serves as the
anode~\cite{nanopattern}, while a platinum electrode serves as the
cathode.  The applied anodizing voltage primarily controls the
diameter and spacing of the pores in the template. The diameter can
be varied from a few nanometers to about 100 nanometers with a pore
depth of a few microns. This technique, therefore, allows for
applying band gap engineering to quantum wires.  Using this
technology, one can create a focal plane array of the detectors
tuned to several wavelengths of IR radiation. One advantage of this
template technique is that fabrication can occur on the final
substrate, which can also support the electronic circuits necessary
for the performance of the detector~\cite{nanopattern}.

In this paper we present initial calculations for a proposed
mechanism~\cite{Crouse_SPIE} for developing such detector arrays.
The structure uses a heterojunction  between two quantum wires
(active and barrier layers). The active layer is n-doped, while the
barrier layer is undoped. In the structure~\cite{Crouse_SPIE}, the
active layer terminates at a metallic contact and the barrier layer
terminates at a transparent conducting oxide, which is exposed to
the incident light. The quantum wires are grown in the template and,
therefore, have the same diameter as the nanopatterned structure.
The small size of the quantum wire results in the quantization of
the conduction band energy levels.

\begin{figure}
  \includegraphics[width=180pt]{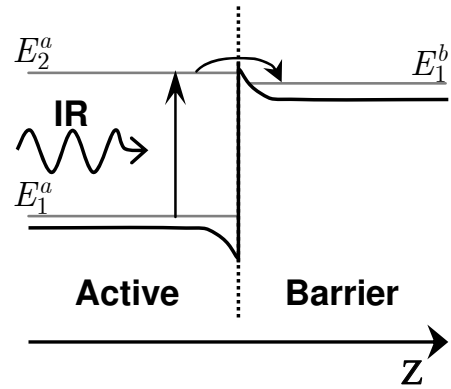}\\
  \vspace{.1in}
  \caption{Conduction band diagram of the IR detector based on a heterojunction between two quantum wires.}
  \label{construction}
\end{figure}

The radii of the wires are selected in such a way that the band
structure is similar to the one shown in Fig.~(\ref{construction}).
The ground state, $E_1^b$, of the barrier layer wire  is slightly
higher than the first excited state, $E_2^a$, of the active layer
wire.  In the absence of IR radiation, only the lowest subband,
$E_1^a$, of the active layer is populated with electrons, which do
not have sufficient energy to overcome the potential barrier across
the junction.  The absorption of IR energy moves some of the
electrons from $E_1^a$ to the next higher subband $E_2^a$. In the
presence of the bias, these higher energy electrons can easily cross
into the barrier layer, thus producing a photocurrent. An accurate
modelling of such a detector and calculation of its characteristics
requires a self-consistent analysis of the carrier kinetics and the
band structure of the system in the vicinity of the junction. Before
that, however, it is important to determine if the suggested concept
is at all possible. In order to answer this question we investigate
the conditions needed to produce a band diagram as shown in
Fig.~(\ref{construction}) by varying radii and chemical compositions
of the active and barrier layers.
\begin{figure}
  \includegraphics[width=250pt]{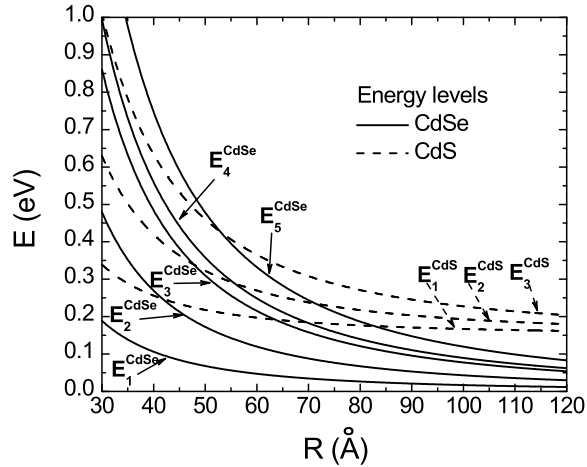}
  \caption{Dependence of the lowest energy states of the conduction electron on the radius of the quantum wire for $CdS$ (dashed line) and $CdSe$ (solid line) for longitudinal wave number
  $k_z$ equal to zero. Energy levels of $CdS$ are shifted up 150 $meV$ due to the conduction band discontinuity.}
  \label{energy_levels}
\end{figure}

Thus, the main task of this paper is to calculate the structure of
subbands in the conductive band of two possible candidates for these
detectors, namely, $CdS$ and $CdSe$. Since we are only interested in
conduction subbands, we can solve a simple single-band problem for
the conduction states using the envelope-function method within the
effective mass approximation. A potential barrier for electrons in
the radial direction of the wire is relatively large ($\sim 2.5
eV$), therefore, it makes sense for initial calculations to restrict
our consideration to the model of the infinite potential barrier. We
will also ignore here the effects of mass and dielectric mismatches.
The Hamiltonian for such a simplified model in the absence of the
bias takes the form
\begin{equation}
\hat{H}=-\frac{\hbar^2}{2m_e}\triangle+V(\rho), \label{Hamiltonian}
\end{equation}
where $m_e$ is the effective mass of electron, and $V(\rho)$ is
defined as
\begin{equation}
V(\rho)=
\begin{cases}
0 & \text{ for } \rho \le R,\\
\infty & \text{ for } \rho > R.
\end{cases}
\end{equation}
Here $R$ is the radius of the wire, which, we assume to have the
shape of a cylinder. Using the cylindrical symmetry of the system
and the homogeneity along $z$, we can find solutions to the
respective Schr\"{o}dinger equation in the form
\begin{equation}\label{wave_function}
\Psi_{n,m}(\rho,\phi,z,k_z)=A_{n,m}J_n(\kappa_{n,m}\rho)e^{im\phi}e^{ik_zz},
\end{equation}
where $\kappa_{n,m}$ are the $m$-th zeroes of the $n$-th order
Bessel function, $J_n(\kappa_{n,m}R)=0$.

The boundary conditions at the infinite potential well require the
wave function to vanish at the boundary. Thus the energy subbands
are determined by zeroes of the Bessel function:
\begin{equation}
E_{n,m}(k_z)=\frac{\hbar^2}{2m_e}\left(\kappa_{n,m}^2+k_z^2\right).\label{Energy}
\end{equation}
The energy levels of the confined electron are counted from the
bottom of the conduction band. Fig.~(\ref{energy_levels}) shows the
positions of the bottoms ($k_z=0$) of several lowest subbands for
$CdS$ ($m_e=0.21m_0$, where $m_0$ is the mass of the free electron)
and $CdSe$ ($m_e=0.13m_0$) as a function of the wire radius.
\begin{figure}
  \includegraphics[width=250pt]{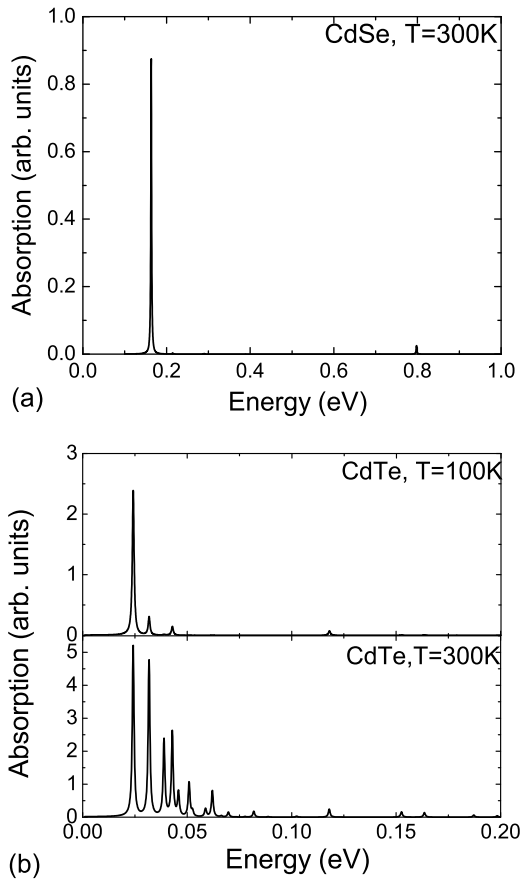}
 \caption{(a) Absorption spectra of intersubband transitions in the conduction band of $CdSe$ quantum wire of radius 40 $\mathrm{\AA}$ calculated
  for temperature 300 K. The $\delta$-function is replaced by a Lorentzian with the coefficient of broadening $\gamma=1$ $meV$.
   (b) The same for the $CdTe$ quantum wire of radius 104 $\mathrm{\AA}$ for temperatures 300 K and 100 K, $\gamma=0.5$ $meV$.}
  \label{AbsorptionCdSe_CdTe}
\end{figure}
The energy in this figure is counted from the bottom of the
conduction band of $CdSe$, which lies 150 $meV$ below the conduction
band of $CdS$\cite{Halsall}. This figure shows that by choosing the
radius of the wire, we can engineer  energy subbands in the active
and barrier wires in a configuration, optimal for detecting light of
a particular frequency. As an example, we consider the $CdSe/CdS$
system. For this detection scheme to work, the first excited subband
in the active $CdSe$ layer should lie above or at the lowest subband
of the barrier $CdS$ material. Fig.~\ref{energy_levels} shows that
this condition may only be fulfilled if the radius of the wire is no
larger than 41.7 $\mathrm{\AA}$. For the $CdSe$ quantum wire of
radius 40 $\mathrm{\AA}$ the ground state energy ($E_1^{CdSe}$) is
equal to 106 $meV$ and the first excited state energy ($E_2^{CdSe}$)
is equal to 270 $meV$. The lowest subband in $CdS$, in this case,
lies slightly below $E_2^{CdSe}$ in $CdSe$. Thus, the structure with
this radius can be used to detect IR photons with the wavelength
7.57 $\mu m$.

Once the band structure is determined, we use the first-order
perturbation theory to calculate absorption spectra of the single
heterostructure:
\begin{equation}
\begin{split}
\alpha(\omega)\sim&\frac{1}{m_e\omega R^2}\sum_{f,i}\int
dk_z\left|\left<f|\hat{H}_{rad}|i\right>\right|^2\times\\
&\delta\left(E_f-E_i-\hbar\omega\right)\left[F(E_i)-F(E_f)\right],
\label{coeff_absropt_initial}
\end{split}
\end{equation}
where $F$ is the  Fermi-Dirac distribution function; $f$ and $i$
represent the set of quantum numbers specifying final and initial
states of the electron. $\hat{H}_{rad}$ here is the interaction
Hamiltonian between the electron and electromagnetic field in the
Coulomb gauge: $\hat{H}_{rad}=\frac{e}{m_e}\bf{A}\cdot\bf{p}$, where
$\bf{A}$ is the vector potential of the electromagnetic field and
$\bf{p}=-i\hbar\nabla$ is the linear momentum operator. Evaluation
of the matrix elements in Eq.~(\ref{coeff_absropt_initial}) reveals
the  selection rule, $n_f=n_i\pm1$, and the necessity for the
incident radiation to be polarized perpendicularly to the axis of
the wire. The momentum of the photon is assumed to be zero.
Substituting Eq.~(\ref{wave_function}) into
Eq.~(\ref{coeff_absropt_initial}) we obtain a final equation for the
absorption coefficient.
\begin{equation}
\begin{split}
&\alpha(\omega)\sim\frac{1}{m_e^2\omega
R^2}\sum_{n=-\infty}^{\infty}\sum_{n'}\sum_{m_f,m_i}A_{n',m_f}A_{n,m_i}\times\\
&\Phi_{n,n'}(R)\left(e^{-E_n/k_BT}-e^{-E_{n'}/k_BT}\right)
\delta\left(E_{n'}-E_n-\hbar\omega\right), \label{final_absorption}\\
\end{split}
\end{equation}
where we have introduced
\begin{equation*}
\begin{split}
\Phi_{n,n'}(R)=&\frac{\kappa_n}{\kappa_{n'}^2-\kappa_n^2}\left(\kappa_{n'}J_{n'}(\kappa_nR)J_{n'+1}(\kappa_{n'}R)-\right.\\
&\left.\kappa_nJ_{n'}(\kappa_{n'}R)J_{n'+1}(\kappa_nR)\right),
\end{split}
\end{equation*}
and $\kappa_{n'}=\kappa_{n',m_f},$ $\kappa_n=\kappa_{n,m_i},$
$E_{n'}=E_{n',m_f},$ $E_n=E_{n,m_i},$ $n'=n\pm1$. Since $E_{f,i}\gg
E_{Fermi}$ and $k_BT\ll E_{f,i}$, we have replaced the Fermi-Dirac
function with the Maxwell-Boltzmann distribution.
Fig.~(\ref{AbsorptionCdSe_CdTe}a) shows the absorption spectra of
intersubband transitions in the conduction band of $CdSe$ of radius
40 $\mathrm{\AA}$ for temperature 300 K. In this figure, the
positions and relative heights of different peaks  at the same
temperature reveal  relative contributions of various transitions to
the spectrum. In the case of $CdSe$ we can see that, at all
temperatures, only the transition from the lowest to the first
excited subband contributes to the spectrum. Other transitions are
strongly suppressed even at higher temperatures because of the small
population of the respective subbands. It should be noted, however,
that when calculating the spectra we introduced a small temperature
independent broadening parameter in order to regularize the
$\delta$-function.

As another example we consider the $CdS/CdTe$ heterojunction, the
band alignment parameters for which can be found in
Ref.~\onlinecite{Fritsche}. Analysis of the subband
 structure for this system shows that in this case  there is an alternative way
for transportation of the electron from the active wire to the
barrier wire. At a radius equal to 74 $\mathrm{\AA}$, the first
excited states in both semiconductors are almost degenerate. This
allows for the electron excited by the 26.03 $\mu m$ photon to the
first excited subband in $CdTe$ to move to $CdS$, occupying the
first excited state in this material. For illustration, we also show
in Fig.~(\ref{AbsorptionCdSe_CdTe}b) the absorption spectra of a
$CdTe$ wire (radius 104 $\mathrm{\AA}$) for temperatures of 100 K
and 300 K. Unlike the case of $CdSe$, in this material increasing
the temperature results in more transitions becoming visible;
however, the main transition corresponding to the 51.69 $\mu m$
photon still remains more intensive at all temperatures.

Our calculations show the feasibility of designing an IR detector
based on intersubband transitions in the junction of two different
semiconductor quantum wires with the same radii. Variations of the
radii and conduction band offsets enables one to determine the
optimal structure required for the detection of a single IR
wavelength. Additionally, focal plane arrays of quantum wires
incorporating different parameters can be used to detect a range of
wavelengths.

The Queens College group acknowledges support from the Air Force
Office of Scientific Research under grant F49620-02-1-0305 and
PSC-CUNY grants, MJP acknowledges support from the Air Force Office
of Scientific Research under grant F49620-01-1-0553 and the Army
Research Office under grant W911NF-04-1-0370.


\end{document}